# A FP-Tree Based Approach for Mining All Strongly Correlated Pairs without Candidate Generation[*]

Zengyou He, Xiaofei Xu, Shengchun Deng

*Department of Computer Science and Engineering Harbin Institute of Technology,*

*92 West Dazhi Street, P.O Box 315, P. R. China, 150001*

Email: zengyouhe@yahoo.com, {xiaofei, dsc}@hit.edu.cn

**Abstract** Given a user-specified minimum correlation threshold $\theta$ and a transaction database, the problem of mining all-strong correlated pairs is to find all item pairs with Pearson's correlation coefficients above the threshold $\theta$. Despite the use of upper bound based pruning technique in the Taper algorithm [1], when the number of items and transactions are very large, candidate pair generation and test is still costly.

To avoid the costly test of a large number of candidate pairs, in this paper, we propose an efficient algorithm, called Tcp, based on the well-known FP-tree data structure, for mining the complete set of all-strong correlated item pairs. Our experimental results on both synthetic and real world datasets show that, Tcp's performance is significantly better than that of the previously developed Taper algorithm over practical ranges of correlation threshold specifications.

## 1 Introduction

More recently, the mining of statistic correlations is applied to transaction databases [1], which retrieves all pairs of items with high positive correlation in a transaction databases. The problem can be formalized as follows [1]. Given a user-specified minimum correlation threshold $\theta$ and a transaction database with $N$ items and $T$ transactions, all-strong pairs correlation query finds all item pairs with correlations above the minimum correlation threshold $\theta$.

Different from the association-rule mining problem [2-4], it is well known that an item pair with high support may have a very low correlation. Additionally, some item pairs with high correlations may also have very low support. Hence, we have to consider all possible item pairs in the mining process. Consequently, when the number of items and transactions are very large, candidate pair generation and test will be very costly.

To efficiently identify all strong correlated pairs, Xiong et al. [1] proposed the Taper algorithm, in which an upper bound of Pearson's correlation coefficient is provided to prune candidate pairs. As shown in [1], the upper bound based pruning technique is very effective in eliminating a large portion of item pairs in the candidate generation phase. However, when the database contains a large number of items and transactions, even testing those remaining candidate

---

[*] This work was supported by the High Technology Research and Development Program of China (No. 2002AA413310, No. 2003AA4Z2170, 2003AA413021) and the IBM SUR Research Fund.

pairs over the whole transaction database is still costly.

Motivated by the above observation, we examine other possible solutions for the problem of correlated pairs mining. We note that, the FP-tree [4] approach does not rely on the candidate generation step. We, therefore, consider how to make use of the FP-tree for the mining of all pairs of items with high positive correlation.

An FP-tree (frequent pattern tree) is a variation of the trie data structure, which is a prefix-tree structure for storing crucial and compressed information about support information. Thus, it is possible to utilize such information to alleviate the multi-scan problem to speed up mining by working directly on the FP-Tree.

In this paper, we propose an efficient algorithm, called Tcp (FP-**T**ree based **C**orrelation **P**airs Mining), based on the FP-tree data structure, for mining the complete set of all strong correlated pairs between items.

The Tcp algorithm consists of two steps: FP-tree construction and correlation mining. In the FP-tree construction step, the Tcp algorithm constructs a FP-tree from the transaction database. In the original FP-tree method [4], the FP-tree is built only with the items with sufficient support. However, in our problem setting, there is no support threshold used initially. We, therefore, propose to build a complete FP-tree with all items in the database. Note that this is equivalent to setting the initial support threshold to zero. The size of an FP-tree is bounded by the size of its corresponding database because each transaction will contribute at most one path to the FP-tree, with the path length equal to the number of items in that transaction. Since there is often a lot of sharing of frequent items among transactions, the size of the tree is usually much smaller than its original database [4]. In the correlation-mining step, we utilize a pattern-growth method to generate all item pairs and compute their *exact* correlation values, since we can get all support information from the FP-tree. That is, different from the Taper algorithm, Tcp algorithm doesn't need to use the upper bound based pruning technique. Hence, Tcp algorithm is insensitive to the input minimum correlation threshold $\theta$, which is highly desirable in real data mining applications. In contrast, the performance of Taper algorithm is highly dependent on both data distribution and minimum correlation threshold $\theta$, because these factors have a great effect on the results of upper bound based pruning.

Our experimental results show that Tcp's performance is significantly better than that of the Taper algorithm for mining correlated pairs on both synthetic and real world datasets over practical ranges of correlation threshold specifications.

## 2 Related Work

Association rule mining is a widely studied topic in data mining research. However, it is well recognized that true correlation relationships among data objects may be missed in the support-based association-mining framework. To overcome this difficulty, correlation has been adopted as an interesting measure since most people are interested in not only association-like co-occurrences but also the possible strong correlations implied by such co-occurrences. Related literature can be categorized by the correlation measures.

Brin et al. [5] introduced *lift* and a $\chi^2$ correlation measure and developed methods for mining such correlations. Ma and Hellerstein [6] proposed a mutually dependent patterns and an

Apriori-based mining algorithm. Recently, Omiecinski [7] introduced two interesting measures, called *all confidence* and *bond*. Both have the downward closure property. Lee et al. [8] proposed two algorithms by extending the pattern-growth methodology [4] for mining *all confidence* and *bond* correlation patterns. As an extension to the concepts proposed in [7], a new notion of the *confidence-closed* correlated patterns is presented in [9]. And an algorithm, called CCMine for mining those patterns is also proposed in [9]. Xiong et al. [10] independently proposed the *h-confidence* measure, which is equivalent to the *all confidence* measure of [7].

In [1], efficiently computing Pearson's correlation coefficient for binary variables is considered. In this paper, we focus on developing a FP-Tree base method for efficiently identifying the complete set of all strong correlated pairs between items, with Pearson's correlation coefficient as correlation measure.

## 3 Background-Pearson's Correlation Coefficient

In statistics, a measure of association is a numerical index that describes the strength or magnitude of a relationship among variables. Relationships among nominal variables can be analyzed with nominal measures of association such as Pearson's Correlation Coefficient. The $\phi$ correlation coefficient [1] is the computation form of Pearson's Correlation Coefficient for binary variables.

As shown in [1], when adopting the support measure of association rule mining [2], for two items A and B in a transaction database, we can derive the support form of the $\phi$ correlation coefficient as shown below in Equation (1):

$$\phi = \frac{\sup(A,B) - \sup(A)*\sup(B)}{\sqrt{\sup(A)*\sup(B)*(1-\sup(A))*(1-\sup(B))}} \qquad (1)$$

where sup(*A*), sup(*B*) and sup(*A*,*B*) are the support of item(s) *A*,*B* and *AB* separately.

Furthermore, as shown in [1], given an item pair {*A*, *B*}, the support value sup(*A*) for item *A*, and the support value sup(*B*) for item B, without loss of generality, let sup(*A*)>=sup(*B*). The upper bound $upper(\phi_{(A,B)})$ of an item pair {*A*,*B* } is:

$$upper(\phi_{(A,B)}) = \sqrt{\frac{\sup(B)}{\sup(A)}} * \sqrt{\frac{1-\sup(A)}{1-\sup(B)}} \qquad (2)$$

As can be seen in Equation (2), the upper bound of $\phi$ correlation coefficient for an item pair {*A*, *B*} only relies on the support value of item *A* and the support value of item *B*. In other words, there is no requirement to get the support value sup(*A*, *B*) of an item pair{*A*, *B*} for the calculation of this upper bound. As a result, in the original Taper algorithm [1], this upper bound is used to serve as a coarse filter to filter out item pairs that are of no interest, thus saving I/O cost by reducing the computation of the support value of those pruned pairs.

The Taper algorithm [1] is a two-step filter-and-refine mining algorithm, which consists of two steps: filtering and refinement. In the filtering step, the Taper algorithm applies the upper bound of the $\phi$ correlation coefficient as a coarse filter. In other words, if the upper bound of the $\phi$ correlation coefficient for an item pair is less than the user-specified correlation threshold, we can prune this item pair right way. In the refinement step, the Taper algorithm computes the exact correlation for each surviving pair from the filtering step and retrieves the pairs with correlations above the user-specified correlation threshold as the mining results.

# 4 Tcp Algorithm

In this section, we introduce Tcp algorithm for mining correlated pairs between items. We adopt ideas of the FP-tree structure [4] in our algorithm.

## 4.1 FP-Tree Basics

An FP-tree (frequent pattern tree) is a variation of the trie data structure, which is a prefix-tree structure for storing crucial and compressed information about frequent patterns. It consists of one root labeled as "NULL, a set of item prefix sub-trees as the children of the root, and a frequent item header table. Each node in the item prefix sub-tree consists of three fields: item-name, count, and node-link, where item-name indicates which item this node represents, count indicates the number of transactions containing items in the portion of the path reaching this node, and node-link links to the next node in the FP-tree carrying the same item-name, or null if there is none. Each entry in the frequent item header table consists of two fields, item name and head of node-link. The latter points to the first node in the FP-tree carrying the item name.

Let us illustrate by an example in [4] the algorithm to build an FP-tree using a user specified minimum support threshold. Suppose we have a transaction database shown in Fig. 1 with minimum support threshold 3.

By scanning the database, we get the sorted (item: support) pairs, (*f*: 4), (*c*:4), (*a*:3), (*b*:3), (*m*:3), (*p*:3) (of course, the frequent 1-itemsets are: *f, c, a, b, m, p*). We use the tree construction algorithm in [4] to build the corresponding FP-tree. We scan each transaction and insert the frequent items (according to the above sorted order) to the tree. First, we insert <*f, c, a, m, p*> to the empty tree. This results in a single path:

root (NULL)→ (*f*: 1)→(*c*:1)→(*a*:1)→(*m*:1)→(*p*:1)

Then, we insert <*f, c, a, b, m*>. This leads to two paths with *f, c* and *a* being the common prefixes:

root (NULL)→ (*f*: 2) →(*c*:2)→(*a*:2)→(*m*:1)→(*p*:1)

and

root (NULL)→ (*f*: 2)→(*c*:2)→(*a*:2)→(*b*:1)→(*m*:1)

Third, we insert <*f,b*>. This time, we get a new path:

root (NULL)→ (*f*: 3)→(*b*:1)

In a similar way, we can insert <*c, b, p*> and <*f, c, a, m, p*> to the tree and get the resulting tree as shown in Fig.1, which also shows the horizontal links for each frequent 1-itemset in dotted-line arrows.

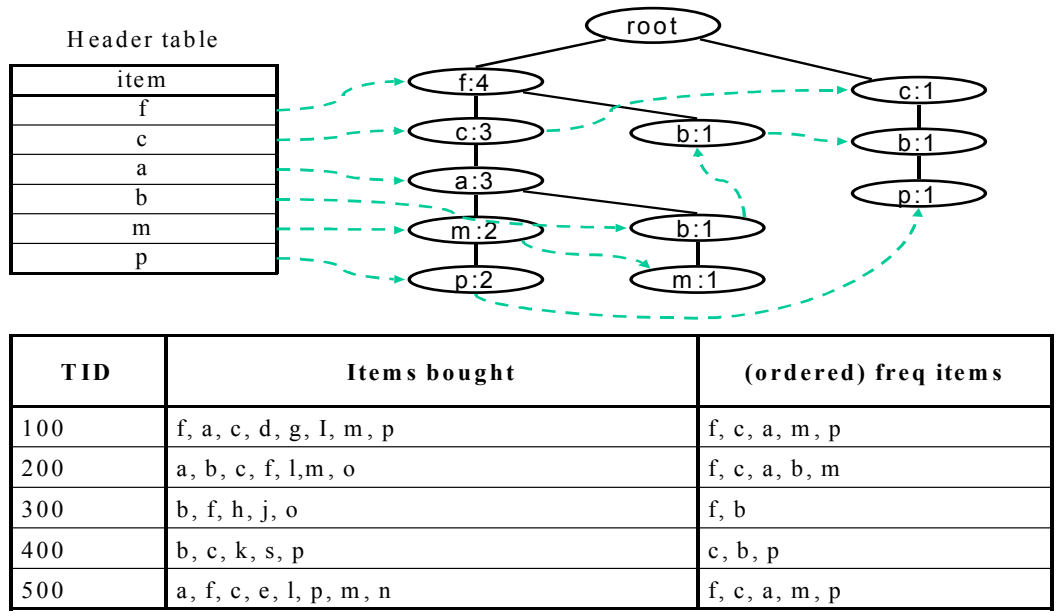

| TID | Items bought | (ordered) freq items |
|-----|--------------|----------------------|
| 100 | f, a, c, d, g, I, m, p | f, c, a, m, p |
| 200 | a, b, c, f, l,m, o | f, c, a, b, m |
| 300 | b, f, h, j, o | f, b |
| 400 | b, c, k, s, p | c, b, p |
| 500 | a, f, c, e, l, p, m, n | f, c, a, m, p |

**Fig. 1**. Database and corresponding FP-tree.

With the initial FP-tree, we can mine frequent itemsets of size $k$, where $k>=2$. An FP-growth algorithm [4] is used for the mining phase. We may start from the bottom of the header table and consider item $p$ first. There are two paths: $(f: 4)\rightarrow(c:3)\rightarrow(a:3)\rightarrow(m:3)\rightarrow(p:2)$ and $(c:1)\rightarrow(b:1)\rightarrow(p:1)$. Hence, we get two prefix paths for $p$: <*fcam*: 2> and <*cb*: 1>, which are called $p$'s conditional pattern base. We also call $p$ the base of this conditional pattern base. An FP-tree on this conditional pattern base (conditional FP-tree) is constructed, which acts as a transaction database with respect to item $p$. Recursively mining this resulting FP-tree to form all the possible combinations of items containing $p$. Similarly, we consider items $m,b,a,c$ and $f$ to get all frequent itemsets.

## 4.2 Tcp Algorithm

The Tcp algorithm consists of two steps: FP-tree construction and correlation mining. In the FP-tree construction step, the Tcp algorithm constructs a FP-tree from the transaction database. In the correlation-mining step, we utilize a pattern-growth method to generate all item pairs and compute their *exact* correlation value. Fig.2 shows the main algorithm of Tcp. In the following, we will illustrate each step thoroughly.

**FP-tree construction step**

In the original FP-tree method [4], the FP-tree is built only with the items with sufficient support. However, in our problem setting, there is no support threshold used initially. We, therefore, propose to build a complete FP-tree with all items in the database (Step 3-4). Note that this is equivalent to setting the initial support threshold to zero. The size of an FP-tree is bounded by the size of its corresponding database because each transaction will contribute at most one path to the FP-tree, with the path length equal to the number of items in that transaction. Since there is often a lot of sharing of frequent items among transactions, the size of the tree is usually much smaller than its original database [4].

As can be seen from Fig.2, before constructing the FP-Tree, we use an additional scan over the database to count the support of each item (Step 1-2). At the first glance, one may argue that this additional I/O cost is not necessary since we have to build the complete FP-Tree with zero support. However, the rationale behind is as follows.

(1) Knowing the support of each item will help us in building the more compact FP-Tree, e.g., ordered FP-Tree.
(2) In the correlation-mining step, single item support is required in the computation of the correlation value for each item pair.

**Tcp Algorithm**

**Input:**
(1): A transaction database $D$
(2): $\theta$ : A user specified correlation threshold

**Output:**
$CP$: All item pairs having correlation coefficients above correlation threshold

**Steps:**
**(A) FP-Tree Construction Step**
  (1) Scan the transaction database, $D$. Find the support of each item
  (2) Sort the items by their supports in descending order, denoted as *sorted-list*
  (3) Create a FP-Tree T, with only a root node with label being "NULL".
  (4) Scan the transaction database again to build the whole FP-Tree with all ordered items
**(B) Correlation Mining Step**
  (5) For each item Ai in the header table of T do
  (6)   Get the conditional pattern base of Ai and compute its conditional FP-Tree Ti
  (7)   For each item Bj in the header table of Ti do
  (8)     Compute the correlation value of (Ai, Bj), $\phi(Ai, Bj)$
  (9)     if $\phi(Ai, Bj) \geq \theta$ then add (Ai, Bj) to $CP$
  (10) Return $CP$

Fig. 2. The Tcp Algorithm

**Correlation mining step**

In the correlation-mining step, we utilize a pattern-growth method to generate all item pairs and compute their *exact* correlation values (Step 5-9), since we can get all support information from the FP-tree.

As can be seen from Fig.2, different from the Taper algorithm, Tcp algorithm doesn't need to use the upper bound based pruning technique, since in Step 8 and Step 9 we can directly compute the correlation values. Therefore, the Tcp algorithm is insensitive to the input minimum correlation threshold $\theta$, which is highly desirable in real data mining applications. In contrast, the performance of Taper algorithm is highly dependent on both data distribution and minimum correlation threshold $\theta$, because these factors have a great effect on the results of upper bound based pruning.

# 5 Experimental Results

A performance study has been conducted to evaluate our method. In this section, we describe those experiments and their results. Both real-life datasets from UCI [12] and synthetic datasets were used to evaluate the performance of our Tcp algorithm again Taper algorithm.

All algorithms were implemented in Java. All experiments were conducted on a Pentium4-2.4G machine with 512 M of RAM and running Windows 2000.

## 5.1 Experimental Datasets

We experimented with both real datasets and synthetic datasets. The mushroom dataset from UCI [12] has 22 attributes and 8124 records. Each record represents physical characteristics of a single mushroom. From the viewpoint of transaction databases, the mushroom dataset has 119 items totally.

The synthesized datasets are created with the data generator in ARMiner software[1], which also follows the basic spirit of well-known IBM synthetic data generator for association rule mining. The data size (i.e., number of transactions), the number of items and the average size of transactions are the major parameters in the synthesized data generation. Table 1 shows the four datasets generated with different parameters and used in the experiments. The main difference between these datasets is that they have different number of items, range from 400 to 1000.

**Table 1**. Test Synthetic Data Sets

| Data Set Name | Number of Transactions | Number of Items | Average Size of Transactions |
|---|---|---|---|
| T10I400D100K | 100,000 | 400 | 10 |
| T10I600D100K | 100,000 | 600 | 10 |
| T10I800D100K | 100,000 | 800 | 10 |
| T10I1000D100K | 100,000 | 1000 | 10 |

## 5.2 Empirical Results

We compared the performance of Tcp with Taper on the 5 datasets by varying correlation threshold from 0.9 to 0.1.

Fig. 3 shows the running time of the two algorithms on the mushroom dataset for $\theta$ ranging from 0.9 to 0.1. We observe that, Tcp's running time remains stable over the whole range of $\theta$. The Tcp's stability on execution time is also observed in the consequent experiments on synthetic datasets. The reason is that Tcp doesn't depend on upper bound based pruning technique, and hence is less affected by threshold parameter. In contrast, the execution time for Taper algorithm increases as the correlation thresholds are decreased. When $\theta$ reaches 0.4, Tcp starts to outperform Taper. Although the performance of Tcp is not good as that of Taper when $\theta$ is larger than 0.4, Tcp at least achieves same level performance as that of Taper on the mushroom dataset. The reason for the Tcp's unsatisfied performance is that, for dataset with fewer items, the advantage of Tcp is not very significant. As can be seen in consequent experiments, Tcp's performance is significantly better than that of the Taper algorithm when the dataset has more items and transactions.

---
[1] http://www.cs.umb.edu/~laur/ARMiner/

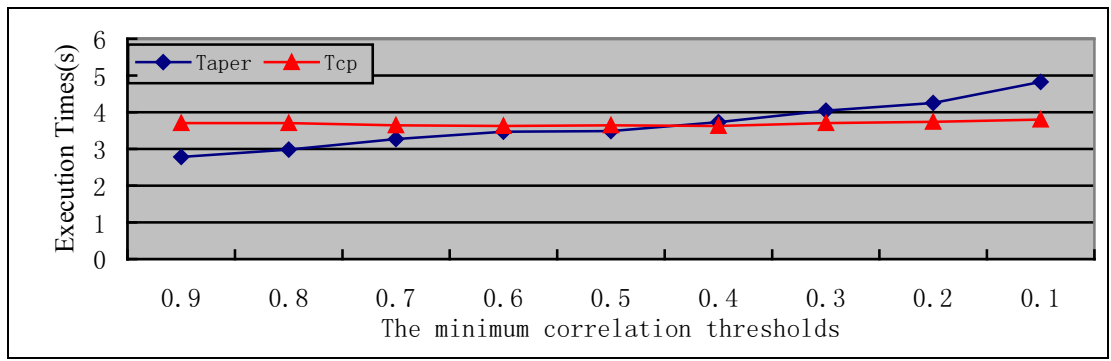

**Fig.3** Execution Time Comparison between Tcp and Taper on Mushroom Dataset

Fig.4, Fig.5, Fig.6 and Fig.7 show the execution times of the two algorithms on four synthetic different datasets as correlation threshold is decreased. From these four figures, some important observations are summarized as follows.

(1) Tcp keeps its stable running time for the whole range of correlation threshold on all the synthetic datasets. It further confirmed the fact that Tcp algorithm is robust with to input parameters.

(2) Taper keeps its increase in running time when the correlation threshold is decreased. And on all datasets, when correlation threshold reaches 0.7, Tcp starts to outperform Taper. Even when the correlation threshold is set to be a very higher value (e.g., 0.9 or 0.8), Tcp's execution time is only a little longer that that of Taper. Besides these extreme large values, Tcp always outperform Taper significantly.

(3) With the increase in the number of items (from Fig.4 to Fig. 7), we can see that the gap between Tcp and Taper on execution time increases. Furthermore, on the T10I1000D100K dataset, when correlation threshold reaches 0.3, the Taper algorithm failed to continue its work because of the limited memory size. While the Tcp algorithm is still very effective on the same dataset even when correlation threshold reaches 0.1. Hence, we can conclude that Tcp is more suitable for mining transaction database with very large number of items and transactions, which is highly desirable in real data mining applications.

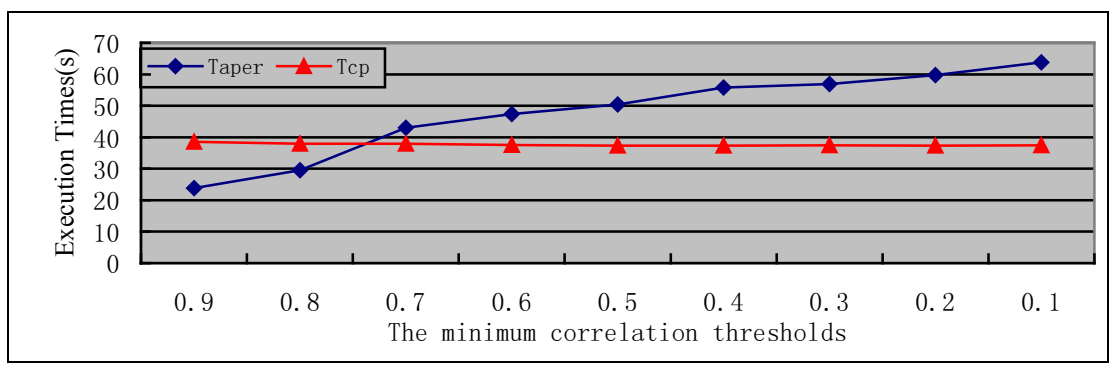

**Fig.4** Execution Time Comparison between Tcp and Taper on T10I400D100K Dataset

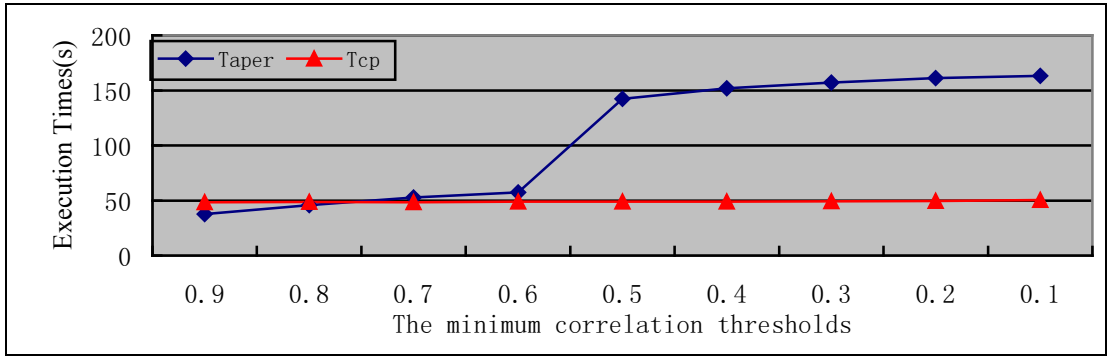

**Fig.5** Execution Time Comparison between Tcp and Taper on T10I600D100K Dataset

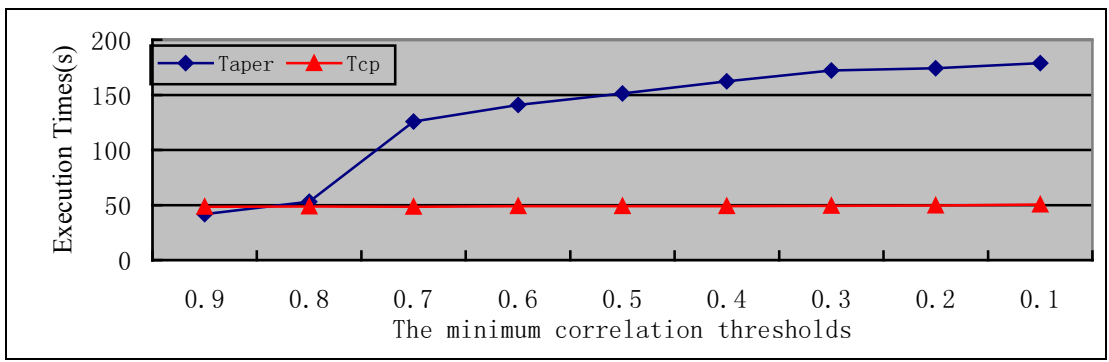

**Fig.6** Execution Time Comparison between Tcp and Taper on T10I800D100K Dataset

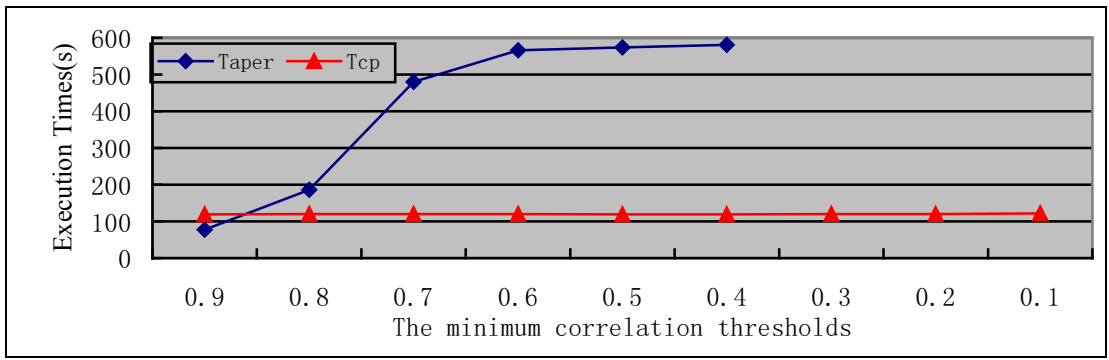

**Fig.7** Execution Time Comparison between Tcp and Taper on T10I1000D100K Dataset

# 6 Conclusions

In this paper, we propose a new algorithm called Tcp for mining statistical correlations on transaction databases. The salient feature of Tcp is that it generates correlated item pairs directly without candidate generation through the usage of FP-Tree structure. We have compared this algorithm to the previously known algorithm, the Taper algorithm, using both synthetic and real world data. As shown in our performance study, the proposed algorithm significantly outperforms the Taper algorithm on transaction database with very large number of items and transactions

In the future, we will work on extending Tcp algorithm for mining Top-*k* correlated pairs, which is more preferable than the minimum correlation-based mining in real applications.

# Reference

[1] H. Xiong, S. Shekhar, P-N. Tan, V. Kumar. Exploiting a Support-based Upper Bound of Pearson's Correlation Coefficient for Efficiently Identifying Strongly Correlated Pairs. In: Proc. of ACM SIGKDD'04, 2004.

[2] R. Agrawal, R. Srikant. Fast Algorithms for Mining Association Rules. In: Proc of VLDB'94, pp. 478-499, 1994.

[3] M. J. Zaki. Scalable Algorithms for Association Mining. IEEE Trans. On Knowledge and Data Engineering, 2000, 12(3): 372~390.

[4] J. Han, J. Pei, J. Yin. Mining Frequent Patterns without Candidate Generation. In: Proc. of SIGMOD'00, pp. 1-12, 2000.

[5] S. Brin, R. Motwani, and C. Silverstein. Beyond market basket: Generalizing association rules to correlations. In Proc. SIGMOD'97, 1997.

[6] S. Ma and J. L. Hellerstein. Mining mutually de- pendent patterns. In Proc. ICDM'01,.2001.

[7] E. Omiecinski. Alternative interest measures for mining associations. IEEE Trans. Knowledge and Data Engineering, 2003.

[8] Y.-K. Lee, W.-Y. Kim, Y. D. Cai, J. Han. CoMine: efficient mining of correlated patterns. In: Proc. ICDM'03, 2003.

[9] W.-Y. Kim, Y.-K. Lee, J. Han. CCMine: Efficient Mining of Confidence-Closed Correlated Patterns. In:Proc. of PAKDD'04, 2004.

[10] H. Xiong, P.-N. Tan, and V. Kumar. Mining hyper-clique patterns with confidence pruning. In Tech. Report, Univ. of Minnesota, Minneapolis, March 2003.

[11] H. T. Reynolds. The Analysis of Cross-classifications. The Free Press, New York, 1977.

[12] C. J. Merz, P. Merphy. UCI Repository of Machine Learning Databases, 1996. ( Http://www.ics.uci.edu/~mlearn/MLRRepository.html).